\documentclass{cpbtex}
\makeatletter
\newcommand{\ie}{\emph{i.e.}\@ifnextchar.{\!\@gobble}{}}
\newcommand{\eg}{\emph{e.g.}\@ifnextchar.{\!\@gobble}{}}
\newcommand{\etc}{etc\@ifnextchar.{}{.\@}}
\makeatother

\usepackage{graphicx} 
\usepackage{caption}  
\usepackage{gensymb}

\begin{document}
\begin{CJK*}{GBK}{song}

\title{Symmetry Quantification and Segmentation in STEM Imaging Through Zernike Moments}


\author{Jiadong Dan$^{1,2}$\thanks{Corresponding author. E-mail:~jiadong.dan@u.nus.edu}, \ Cheng Zhang$^{3}$, \ Xiaoxu Zhao$^4$, and \ N. Duane Loh$^{1,2,3}$\thanks{Corresponding author. E-mail:~duaneloh@nus.edu.sg}\\
$^{1}${Department of Biological Sciences, National University of Singapore}\\  
$^{2}${Center for Bioimaging Sciences (CBIS),  National University of Singapore}\\ 
$^{3}${Department of Physics, National University of Singapore}\\
$^{4}${School of Materials Science and Engineering, Peking University}\\
}   

\date{\today}
\maketitle

\begin{abstract}
We present a method using Zernike moments for quantifying rotational and reflectional symmetries in scanning transmission electron microscopy (STEM) images, aimed at improving structural analysis of materials at the atomic scale. This technique is effective against common imaging noises and is potentially suited for low-dose imaging and identifying quantum defects. We showcase its utility in the unsupervised segmentation of polytypes in a twisted bilayer TaS\textsubscript{2}, enabling accurate differentiation of structural phases and monitoring transitions caused by electron beam effects. This approach enhances the analysis of structural variations in crystalline materials, marking a notable advancement in the characterization of structures in materials science.
\end{abstract}



\section{Introduction}

There are innumerable arrangements of atoms in a crystal. However, this multitude of atomic arrangements can be described and classified by their symmetry features. In 1892, the seminal works by Fedorov and Sch{\"o}nflies proved that there were precisely 230 space groups\ucite{1,2}. This monumental discovery transitioned the field from grappling with seemingly infinite configurations to navigating a finite, manageable set. Deeply rooted in the principles of symmetry, this also underpins the essence of modern crystallography, showcasing symmetry's profound capacity to decipher and elucidate the complexities of the natural world.

Symmetry of structures is critical in materials sciences to understand and predict material properties. Physical properties of many condensed matter systems are governed by symmetry breaking phenomena such as ferroelectricity\ucite{3,4,5}, piezoelectricity\ucite{6,7}, charge density wave (CDW)\ucite{8,9}, and superconductivity\ucite{10,11}. In CDW systems, for example, some metallic transition metal dichalcogenides (TMDs) exhibit commensurate charge-ordering at a low temperature due to periodic lattice distortion (PLD). Specifically, 1T-TaS\textsubscript{2} undergoes a metal-to-insulator transition as the spatial modulation of lattice distortion becomes commensurate with the crystal lattice to form a $\sqrt{13}\times\sqrt{13}$ supercell\ucite{12}. For piezoelectric materials, a giant piezoelectric coefficient ($d_{33}$) can be obtained in centrosymmetric oxides by applying an electric field that breaks the inversion symmetry\ucite{5,7}. The techniques for characterizing and observing these symmetry-related phenomena span diffraction methodologies to real-space imaging, providing a comprehensive view of symmetry's influence on material properties. 

Recent advances in scanning transmission electron microscopy (STEM) have achieved spatial resolution that is smaller than the thermal fluctuation of atoms\ucite{13}. Instrumental development also has improved the temporal resolution - 25 frames per second (fps) STEM imaging with the image size of $512\times512$ pixels\ucite{14}. Progress of spatial-temporal resolution enables dynamical observation of symmetry breaking phenomena in real space\ucite{15}. Moreover, current state-of-the-art STEM can capture vast amounts of data\ucite{16}, presenting unique opportunities for machine learning and data analytics in this field.

One of the central challenges is how to extract and interpret structures with rich symmetry information from large volumes of data. Many machine learning models have been proposed to extract and identify structure symmetry information for image segmentation\ucite{17,18}, defect classification and motif hierarchy constructions\ucite{19,20}. Although these high-throughput models and methods have accelerated material characterization, their breadth of applicability is still limited. A major obstacle comes from the scarcity of experiment data and the diversity of structures in the real world - it is impossible to capture or simulate all structures in nature. Inspired by the reduction of crystals to a finite number of groups, we feel symmetry features are the key solution to reduce the complexity of analyzing vast and varied structural data if we can group STEM images of different materials to a similar symmetry pattern that measures the loss of local symmetry captured by some symmetry scores.

Traditional methods of analyzing local symmetry rely on comparing the data patches before and after symmetry operations (\eg, translation and rotation), which have been investigated in reciprocal and real space imaging\ucite{17,21,22}. The symmetry scores in these studies provide excellent interpretation, but the process for symmetry operation for all local image patches is computationally expensive. Alternatively, variational autoencoders (VAE) can be used to explore order parameters and symmetry breaking in four-dimensional STEM data\ucite{23,24}. The results of the latent space learned by VAE are not readily human-interpretable., which limits their applicability. A fast and intuitive symmetry extraction framework for STEM is needed to automatically analyze the high-level symmetry features present in all materials in a human-interpretable manner.

Here we developed a framework for quantifying rotational and reflectional symmetries in atomic resolution STEM images through Zernike moments, providing a notable advance in the structural analysis of materials at the atomic scale. This approach is robust against common types of imaging noise, offering advantages for low-dose imaging of sensitive materials and the identification of structural defects. We applied this method to accurately segment and differentiate structural phases within a twisted bilayer TaS\textsubscript{2} to demonstrate its practical application, which also enables the monitoring of phase transitions induced by electron beam irradiation. The symmetry scores are designed to be sensitive to deviations from crystallinity, thus enhancing the capability to analyze structural variations in crystalline materials.

%

\begin{figure}[!h]
    \includegraphics[width=\textwidth]{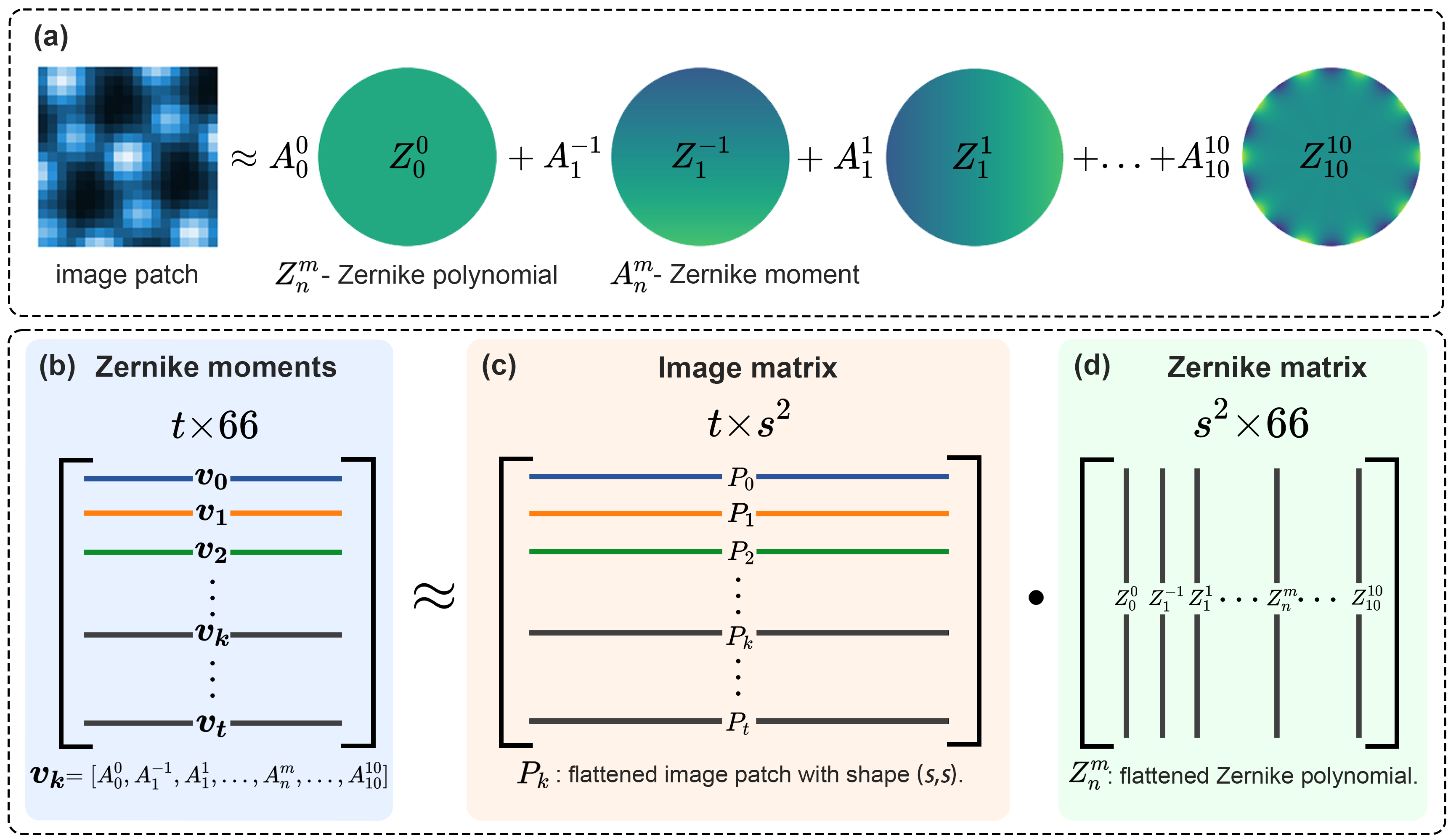}
    \captionsetup{font=small}
    \caption{Representation of image patches as Zernike moments. ($\textbf{a}$) depicts the approximation of an image patch (left) as a linear combination of Zernike polynomials $Z_n^m$ with coefficient $A_n^m$, which is called Zernike moment. Zernike moments can be calculated via dot product, as shown from ($\textbf{b}$) to ($\textbf{d}$).}
    \label{fig:Zernike representation}
\end{figure}


\section{Representation of image patches as Zernike moments}


Zernike polynomials, represented as $Z_n^m$, are an orthogonal set defined on the unit disk, and they are further described in Ref. \cite{19}. These polynomials can be used to represent local STEM image patches, facilitating the identification of structural defects. Given an image patch, $P_k$ can be approximated by a linear combination of Zernike polynomials $Z_n^m$ with corresponding coefficients $A_n^m$, as illustrated in Fig. \hyperref[fig:Zernike representation]{\ref{fig:Zernike representation}(a)}. The Zernike moment $A_{n}^{m}$ can be computed by integrating the elementwise product $P_k\cdot Z_n^m$, which is given by Eq. (\ref{eq:Anm}). Typically, the first 66 polynomials (from $Z_0^0$ to $Z_{10}^{10}$) are employed to encapsulate the shape, contrast, and symmetry characteristics of a local image patch, offering robustness against noise, rotation, and scaling.

\begin{align}\label{eq:Anm}
\begin{split}
A_n^m = \iint P_k \cdot Z_n^m \space \mathrm{d}x \mathrm{d}y
\end{split}
\end{align}

When the shape of the image patch aligns with that of the Zernike polynomial circular patch, the computation of Zernike moments for a collection of images can be efficiently achieved through matrix multiplication, as illustrated in Fig. \hyperref[fig:Zernike representation]{\ref{fig:Zernike representation}(b)}. For image patches $[P_0, P_1, \cdots, P_k, \cdots, P_t]$ each sized with a shape of $(s, s)$, we reshape each patch into a one-dimensional row vector, $P_k$, forming an image matrix. Likewise, each Zernike polynomial can be flattened to form a Zernike matrix consisting of 66 polynomials. The dot product of the image matrix and Zernike matrix yields the Zernike moments $[\bm{v}_0, \bm{v}_1, \cdots, \bm{v}_k, \cdots, \bm{v}_t]$. 

For larger image sizes, exceeding the Zernike polynomial patch, we use a fast two-dimensional convolution method to approximate the Zernike moments. In this approach, 66 Zernike polynomial patches serve as convolution kernels, and the convolution output, once normalized, provides the corresponding Zernike moment. Implementation details are available in our GitHub repository, \href{https://github.com/jiadongdan/motif-learn}{motif-learn} (see \hyperref[code-availability]{Code availability}).

%
\begin{figure}[!h]
    \includegraphics[width=\textwidth]{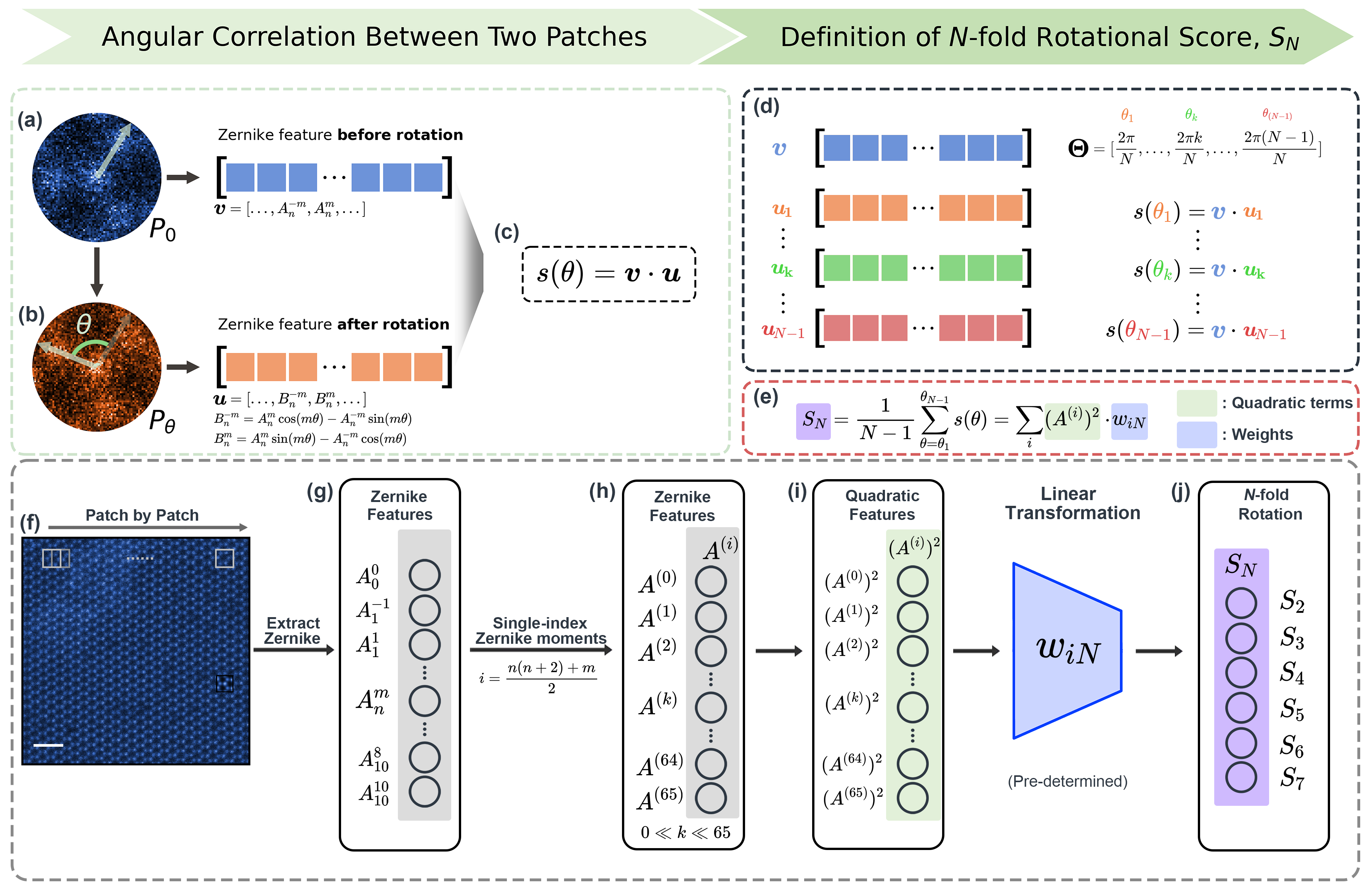}
    \captionsetup{font=small}
    \caption{Definition of \textit{N}-fold rotational symmetry score $S_N$. ($\textbf{a}$) The original HAADF-STEM image patch, denoted as $P_0$, is represented by a Zernike feature vector $\bm{v}=\left[A_0^0,\cdots, A_n^{-m},\cdots, A_n^m,\cdots\right]$. ($\textbf{b}$) The transformed image patch obtained by rotating $P_0$ by an angle $\theta$ anticlockwise, denoted as $P_{\theta}$, is represented by a Zernike features vector $\bm{u}=\left[B_0^0,\cdots, B_n^{-m},\cdots, B_n^m,\cdots\right]$. ($\textbf{c}$) The similarity score of $P_0$ and $P_{\theta}$ is measured by the dot product of $\bm{v}$ and $\bm{u}$, denoted as $s(\theta)$. ($\textbf{d}$) For discrete angles at which the image with perfect \textit{N}-fold rotational symmetry can be rotated while maintaining its orientation, $\bm{\Theta}=\left[ \frac{2\pi}{N}, \dots, \frac{2\pi k}{N},\dots,\frac{2\pi (N-1)}{N} \right]$, a series of similarity scores can be obtained, $\left\{s\left(\theta_{k}\right)\right\}$, where $\theta_{k}=\frac{2\pi k}{N}$, $1\leq k\leq N-1$.  ($\textbf{e}$) The \textit{N}-fold rotational symmetry score ($S_N$) is defined as the mean of these similarity scores, $\left\{s\left(\theta_{k}\right)\right\}$. ($\textbf{f}$) A sliding window technique is used to obtain a series of local image patches. ($\textbf{g}$) The first 66 Zernike moments ranging from $A_{0}^{0}$ to $A_{10}^{10}$ are computed for each patch. ($\textbf{h}$) Zernike moments ($A_n^{m}$) are converted into a single index format ($A^{(i)}$). ($\textbf{i}$) Squaring $A^{(i)}$ to yield the quadratic term of each Zernike feature. ($\textbf{j}$) The \textit{N}-fold rotation symmetry score ($S_N$) is computed as the linear combination of quadratic terms weighted by pre-determined coefficients $w_{iN}$. (scale bar: 1 nm)}
    \label{fig:Fig1}
\end{figure}

\section{Definition of \textit{N}-fold rotational symmetry score $S_N$}

\begin{figure}[!h]
    \includegraphics[width=\textwidth]{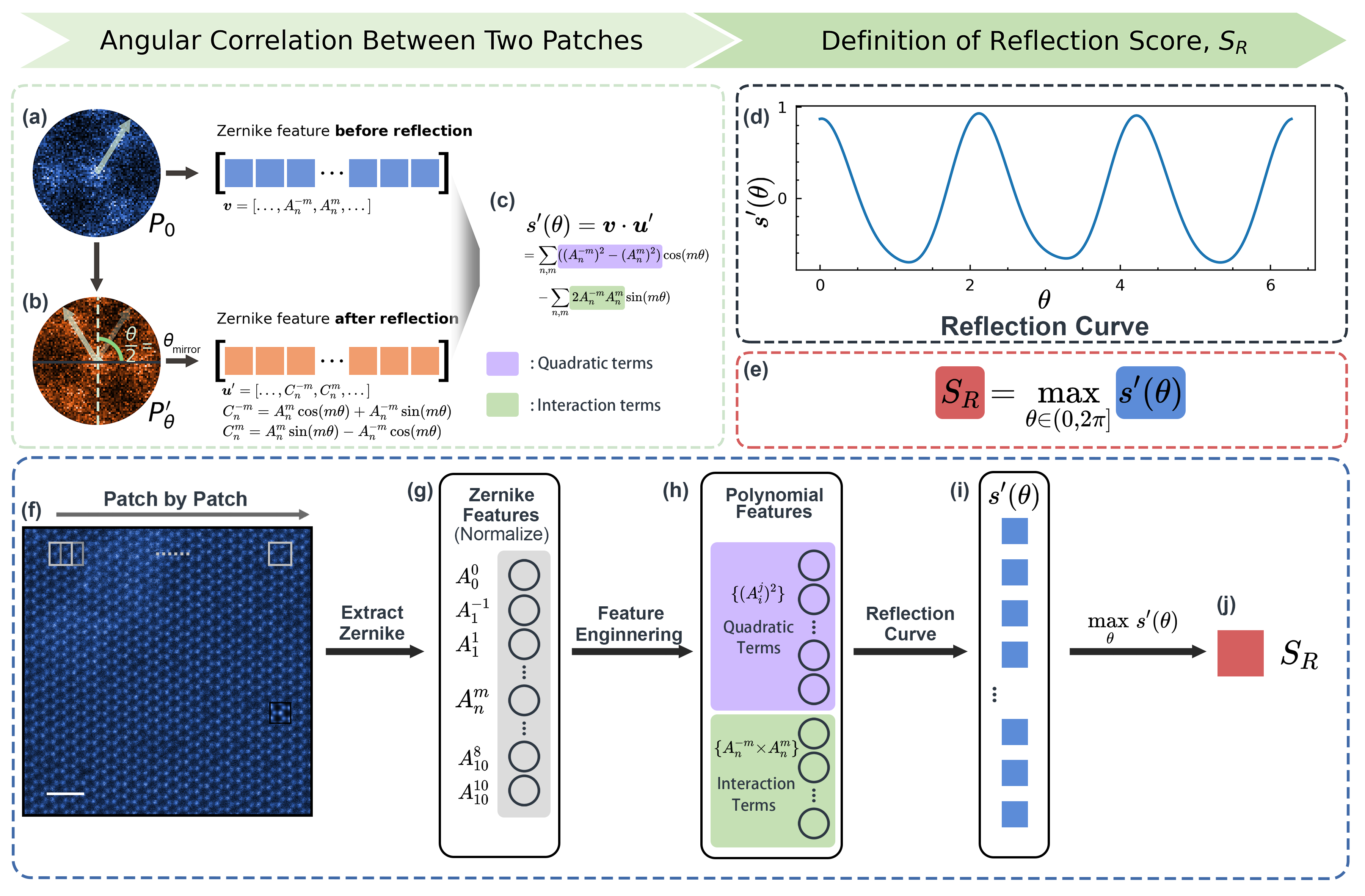}
    \captionsetup{font=small}
    \caption{Definition of reflectional symmetry score $S_R$. ($\textbf{a}$) The original HAADF-STEM image patch, denoted as $P_0$, is represented by a Zernike feature vector $\bm{v}=\left[A_0^0,\cdots, A_n^{-m},\cdots, A_n^m,\cdots\right]$. ($\textbf{b}$) The image patch $P_0$, when reflected across a mirror plane at an angle $\theta_{mirror}=\frac{\theta}{2}$ relative to the horizontal axis, produces the transformed patch denoted by $P_{\theta}^\prime$. This patch is characterized by a Zernike feature vector $\bm{u^\prime}=\left[C_0^0,\cdots, C_n^{-m},\cdots, C_n^m, \ldots\right]$. ($\textbf{c}$) The dot product of $\bm{v}$ and $\bm{u^{\prime}}$, $s^\prime\left(\theta\right)$ measures the similarity score of $P_0$ and $P_{\theta}^{\prime}$, which is the sum of a linear combination of quadratic terms with coefficients $\cos{\left(m\theta\right)}$ and interaction terms with coefficient $-2\sin{\left( m\theta\right)}$. ($\textbf{d}$) For continuous angle $\theta$ between $0$ to $2\pi$, the reflectional curve $s^\prime \left(\theta\right)$ can be obtained by sampling enough number of angles. ($\textbf{e}$) The reflectional score, denoted as $S_R$, is defined as the maximum value of the reflection curve $s^\prime \left(\theta\right)$. ($\textbf{f}$) A sliding window technique is used to obtain a series of local image patches. ($\textbf{g}$) The first 66 Zernike moments ranging from $A_{0}^{0}$ to $A_{10}^{10}$ are computed for each patch. ($\textbf{h}$) The quadratic terms and interaction terms are used to calculate the reflectional curve $s^\prime \left(\theta\right)$. ($\textbf{i}$) Reflection curve $s^\prime\left(\theta\right)$ is visualized by a dense vector. ($\textbf{j}$) Reflectional symmetry score $S_R$ is the maximum of $s^\prime\left(\theta\right)$. (scale bar: 1 nm)}
    \label{fig:Fig2}
\end{figure}

We first define how we compute the \textit{N}-fold rotational symmetry score using Zernike moments. The \textit{N}-fold rotational symmetry score quantifies the degree of resemblance between the object and its rotated versions. Rather than directly comparing image patches, we measure the similarity between an original image patch and its rotated version by computing the dot product of Zernike moments of these image pairs. As illustrated in Fig. \hyperref[fig:Fig1]{\ref{fig:Fig1}(a)}, we can represent an image patch $P_0$ with a Zernike feature $\bm{v} =\left[A_0^0,\cdots, A_n^{-m},\cdots, A_n^m,\cdots\right]$, where $A_n^m$ denotes the Zernike moment with a radial index of $n$ and an azimuthal index of $m$. Similarly, in Fig. \hyperref[fig:Fig1]{\ref{fig:Fig1}(b)}, the image patch rotated by an angle $\theta$, denoted as $P_\theta$, can be represented by another Zernike feature $\bm{u}=\left[B_0^0,\cdots, B_n^{-m},\cdots, B_n^m,\cdots\right]$. The Zernike moments $B_n^{-m}$ and $B_n^m$ are related to $A_n^{-m}$, $A_n^m$, and $\theta$ through the following Eq. (\ref{eq:Bnm}), which are also illustrated in Fig. \hyperref[fig:Fig2]{\ref{fig:Fig2}(b)}.

\begin{eqnarray} \label{eq:Bnm}
\left\{\begin{array}{l}
B_n^{-m}=A_n^m \cos (m \theta)-A_n^{-m} \sin (m \theta) \\
B_n^m=A_n^m \sin (m \theta)+A_n^{-m} \cos (m \theta)
\end{array}\right.
\end{eqnarray}

Without implementing the rotation in real space, we can evaluate the angular correlation of $P_0$ and $P_\theta$ using $\bm{v}$ and $\bm{u}$ only. As depicted in Fig. \hyperref[fig:Fig1]{\ref{fig:Fig1}(c)}, the value $s\left(\theta\right)$, which is the dot product of $\bm{v}$ and $\bm{u}$, represents the cosine similarity of $\bm{v}$ and $\bm{u}$. In mathematical terms, $s\left(\theta\right)$ can be simplified as a sum of squares of Zernike moments (\ie, $\left(A_n^{-m}\right)^2$ or $\left(A_n^{m}\right)^2$) multiplied by the corresponding cosine term $\cos{\left(m\theta\right)}$, as detailed in Eq. (\ref{eq:s(theta)}).

\begin{align}\label{eq:s(theta)}
\begin{split}
s\left(\theta\right) &= \bm{v} \cdot \bm{u} \\
&=\sum_{n,m} \left( \left(A_n^{-m}\right)^2+\left(A_n^m\right)^2\right) \cos{\left(m\theta\right)}
\end{split}
\end{align}

For an image patch that exhibits perfect \textit{N}-fold rotational symmetry, the similarity scores between any pair of images within the same rotational group are expected to be one. This principle applies specifically to a set of equally spaced angles, denoted by $\bm{\Theta}=\left[ \frac{2\pi}{N} , \dots, \frac{2\pi k}{N},\dots,\frac{2\pi (N-1)}{N} \right]$. Here $\bm{\Theta}$ refers to a series of specific angles at which the image with perfect \textit{N}-fold rotational symmetry can be rotated while maintaining its orientation. Thus, for these angles, the similarity score $s\left(\frac{2\pi k}{N}\right)=1$, where $1\le k\le \mathit{N}-1$. To calculate the \textit{N}-fold rotational symmetry score, $S_{N}$, for any given image patch, we take the average of similarity scores between the Zernike feature $\bm{v}$ and the Zernike features of $N-1$ images rotated by these discrete angles, $\bm{u_1}$ to $\bm{u_{N-1}}$, as illustrated in Figs. \hyperref[fig:Fig1]{\ref{fig:Fig1}(d)} and \hyperref[fig:Fig1]{\ref{fig:Fig1}(e)}. Essentially, $S_{\mathit{N}}$ is computed as $\sum_{k=1}^{\mathit{N}-1} s\left(\frac{2\pi k}{\mathit{N}}\right)$. Further analysis indicates that $S_{\mathit{N}}$ can be seen as a linear combination of squares of Zernike moments, with coefficients $w_{iN}$, as outlined in Eq. (\ref{eq:SN}), where $i$ is a single index for Zernike moments $A_{n}^{m}$. A mapping of two indices $m$ and $n$ is detailed in Eq. (\ref{eq:mn2i}).

\begin{align}\label{eq:SN}
\begin{split}
S_{N} &= \frac{1}{\mathit{N}-1} \sum_{\theta = \theta_{1}}^{\theta_{\mathit{N}-1}} s\left(\theta\right)\\
&= \sum_i \left(A^{\left(i\right)}\right)^2 \cdot w_{iN}
\end{split}
\end{align}

\begin{align}\label{eq:mn2i}
\begin{split}
i &= \frac{n(n+2)+m}{2}
\end{split}
\end{align}

\begin{align}\label{eq:wiN}
\begin{split}
w_{iN} &= \frac{1}{N-1} \cdot \sum_{k = 1}^{N-1} \cos{\left( m \frac{2 \pi k}{N}  \right)}
\end{split}
\end{align}

The coefficient $w_{iN}$ is the mean of several cosine terms, and its value only depends on $m$ and $N$, as shown in Eq. (\ref{eq:wiN}).  For each specific value of $m$ and \textit{N}, $w_{iN}$ can be precalculated. As shown in Appendix \hyperref[appen:A3]{A3}, the coefficient is characterized by three distinct cases: $m \leq 1$; $m \geq 2 \land m\pmod{N} = 0$; and $m \geq 2 \land m\pmod{N} \neq 0$. For case (1), Zernike moments with $m=0$ and $m=1$,  we manually set the coefficient to zero; for case (2), Zernike moments characterized by $m\geq2 \land m\pmod{N} = 0$ have a coefficient of $1$; for case (3), Zernike moments where $m\geq2 \land m\pmod{N} \neq 0$, the coefficient is $-\frac{1}{\mathit{N}-1 }$, as shown in Eq. (\ref{eq:wiNcases}). 

\begin{equation}\label{eq:wiNcases}
\begin{split}
w_{i N}= \begin{cases}0, &\text{ when } \quad m \leq 1 \\
1, \quad &\text { when } \quad m \geq 2 \land m\pmod{N}=0 \\
-\frac{1}{N-1},  &\text { when } \quad m \geq 2 \land m\pmod{N} \neq 0\end{cases}
\end{split}
\end{equation}

The tabulated values for these coefficients are also provided in Table 1.

%
\renewcommand\arraystretch{1}
\begin{center}
{\footnotesize{\bf Table 1.}} \textit{N}-fold rotational symmetry coefficients tabulated by  $m$ and \textit{N}.\\
\vspace{2mm}

\begin{tabular}{l| c c c c c c}
\hline
      & \textit{N} = 2 & \textit{N} = 3 & \textit{N} = 4 & \textit{N} = 5 & \textit{N} = 6 & \textit{N} = 7\\
\hline
 $m=0$ & 0 & 0 & 0 & 0 & 0 & 0 \\

 $m=1$ & 0 & 0 & 0 & 0 & 0 & 0 \\

 $m=2$ & 1 & -$\frac{1}{2}$ & -$\frac{1}{3}$ & -$\frac{1}{4}$ & -$\frac{1}{5}$ & -$\frac{1}{6}$ \\

 $m=3$ & -1 & 1 & -$\frac{1}{3}$ & -$\frac{1}{4}$ & -$\frac{1}{5}$ & -$\frac{1}{6}$ \\

 $m=4$ & 1 & -$\frac{1}{2}$ & 1 & -$\frac{1}{4}$ & -$\frac{1}{5}$ & -$\frac{1}{6}$ \\

 $m=5$ & -1 & -$\frac{1}{2}$ & -$\frac{1}{3}$ & 1 & -$\frac{1}{5}$ & -$\frac{1}{6}$ \\

 $m=6$ & 1 & 1 & -$\frac{1}{3}$ & -$\frac{1}{4}$ & 1 & -$\frac{1}{6}$ \\

 $m=7$ & -1 & -$\frac{1}{2}$ & -$\frac{1}{3}$ & -$\frac{1}{4}$ & -$\frac{1}{5}$ & 1 \\

 $m=8$ & 1 & -$\frac{1}{2}$ & 1 & -$\frac{1}{4}$ & -$\frac{1}{5}$ & -$\frac{1}{6}$ \\

 $m=9$ & -1 & 1 & -$\frac{1}{3}$ & -$\frac{1}{4}$ & -$\frac{1}{5}$ & -$\frac{1}{6}$ \\

 $m=10$ & 1 & -$\frac{1}{2}$ & -$\frac{1}{3}$ & 1 & -$\frac{1}{5}$ & -$\frac{1}{6}$ \\
\hline
\end{tabular}
\end{center}

The process for calculating the rotational symmetry score in an experimental high-angle annular dark-field (HAADF) STEM image is illustrated from Fig \hyperref[fig:Fig1]{\ref{fig:Fig1}(f)} to \hyperref[fig:Fig1]{\ref{fig:Fig1}(j)}. We employ a sliding window technique to obtain a series of local image patches as illustrated in Fig. \hyperref[fig:Fig1]{\ref{fig:Fig1}(f)}. By default, we compute the first 66 Zernike moments for each patch within the image, as illustrated in Fig \hyperref[fig:Fig1]{\ref{fig:Fig1}(g)}. These moments ($A_n^{m}$) are then streamlined into a single index format, $A^{(i)}$. Following this step, Fig \hyperref[fig:Fig1]{\ref{fig:Fig1}(i)} demonstrates the calculation of quadratic terms, $(A^{(i)})^2$, which are subsequently weighted by $w_{iN}$ to ascertain the \textit{N}-fold rotational symmetry score ($S_{N}$).

Employing Zernike features to depict image patches bypasses the necessity for labor-intensive transformations of these patches, streamlining the process. Additionally, this approach compacts sizable image patches into a concise set of Zernike moments, saving memory for downstream analysis. Most notably, due to the rotational invariance of Zernike moments, we only need to compute the Zernike moments for an original image patch just once, the Zernike moments for rotated images can be obtained by simple linear transformation, thus reducing computational overhead.

%

\section{Definition of reflectional symmetry score $S_R$}

Similar to the computation of the similarity score of $\bm{v}$ and $\bm{u}$, we first compute the similarity score between an image patch and its reflected version using Zernike moments. Figures \hyperref[fig:Fig2]{\ref{fig:Fig2}(a)} and \hyperref[fig:Fig2]{\ref{fig:Fig2}(b)}  display the original image patch $P_0$ and the reflected image patch $P_{\theta}^{\prime}$, obtained by reflecting $P_0$ along a mirror plane (dash line in Fig. \hyperref[fig:Fig2]{\ref{fig:Fig2}(b)}) at an angle of $\theta_{mirror}=\frac{\theta}{2}$ relative to the horizontal axis. These patches, $P_0$ and $P_{\theta}^{\prime}$, can be represented using Zernike features as vectors $\bm{v} = \left[A_0^0,\cdots, A_n^{-m},\cdots, A_n^m,\cdots \right]$ and $\bm{u}^{\prime} = \left[C_0^0,\cdots, C_n^{-m},\cdots, C_n^m,\cdots \right]$ respectively. The Zernike moments $C_n^{-m}$ and $C_n^m$ are described by Eq. (\ref{eq:Cnm}).

\begin{eqnarray}\label{eq:Cnm}
\left\{\begin{array}{l}
C_n^{-m}=A_n^m \cos (m \theta)+A_n^{-m} \sin (m \theta) \\
C_n^m=A_n^m \sin (m \theta)-A_n^{-m} \cos (m \theta)
\end{array}\right.
\end{eqnarray}

We can define the dot product of $\bm{v}$ and $\bm{u}^{\prime}$ as a similarity score between $P_0$ and $P_\theta^{\prime}$. This similarity is denoted as $s^{\prime}\left(\theta\right)$ and is illustrated in Eq. (\ref{eq:s'(theta)}), also highlighted in Fig. \hyperref[fig:Fig2]{\ref{fig:Fig2}(c)}.

\begin{align}\label{eq:s'(theta)}
\begin{split}
s^{\prime}\left(\theta\right) &= \bm{u} \cdot \bm{v}^{\prime} \\
&=\sum_{n,m} \left( \left(A_n^{-m}\right)^2-\left(A_n^m\right)^2\right) \cos{\left(m\theta\right)}-\sum_{n,m} 2 A_n^{-m} A_n^m\sin{\left(m\theta \right)}
\end{split}
\end{align}
%
%
It is worth noting that, aside from the quadratic terms $ \left(A_n^m\right)^2$ and  $ \left(A_n^{-m}\right)^2$, $s^{\prime}\left(\theta\right)$ now includes additional interaction terms of $A_n^{-m}A_n^m$ compared with Eq. (\ref{eq:s(theta)}). The interaction terms are also highlighted in light green color in Fig. \hyperref[fig:Fig2]{\ref{fig:Fig2}(c)}, which are weighted by $-2\sin{\left(m\theta\right)}$. It is noted that we manually set Zernike moments with $m = 0$ and $m = 1$ to zero, as $A_n^{0}$ only captures the radial symmetry and $A_n^{-1}$ and $A_n^{1}$ do not directly contribute to reflectional symmetry (see Appendix \hyperref[appen:A3]{A3}). We call $s^{\prime}\left(\theta\right)$ the reflectional curve, which forms a continuous curve encompassing all angles within $\theta \in \left(0,\ 2\pi\right]$, corresponding to $\theta_{mirror}\in \left(0,\ \pi\right]$. By sampling a sufficient number of angles, we can obtain $s^{\prime}\left(\theta\right)$ as a dense vector, as depicted in Fig. \hyperref[fig:Fig2]{\ref{fig:Fig2}(d)}. The angles corresponding to the peaks in $s^{\prime}\left(\theta\right)$ indicate the orientations of possible mirror planes. Instead of utilizing a set of discrete angles ($\bm{\Theta}$) to define the rotational symmetry score $S_{N}$, the reflectional symmetry score $S_R$ is defined as the maximum value of $s^{\prime}\left(\theta\right)$, as illustrated in Fig. \hyperref[fig:Fig2]{\ref{fig:Fig2}(e)}.

Figures \hyperref[fig:Fig2]{\ref{fig:Fig2}(f)} to \hyperref[fig:Fig2]{\ref{fig:Fig2}(j)} show the process of calculating the reflectional symmetry score for each pixel within an experimental HAADF-STEM image, with an exclusion mask applied to pixels located within half a patch size from the image boundaries to avoid edge effects. Initially, for each image patch centered on the target pixel, we compute the first 66 Zernike moments (up to $m=10$). Following this, we create quadratic and interaction terms, which are then weighted by coefficients $\cos{\left(m\theta\right)}$ and $\sin{\left(m\theta\right)}$ to generate the reflection curve $s^{\prime}\left(\theta\right)$. Numerically, the reflectional symmetry score is the maximum value of the dense vector, as depicted from Fig. \hyperref[fig:Fig2]{\ref{fig:Fig2}(i)} to Fig. \hyperref[fig:Fig2]{\ref{fig:Fig2}(j)}. 

Algorithmically, the computation process is executed efficiently through kernel convolution and tensor dot product. For large-image size, we use sliding convolutional kernels comprising 66 Zernike polynomials to generate a stack of Zernike moments (in a three-dimensional tensor format). The reflectional symmetry map,  $S_R$, is obtained via a tensor dot product with the coefficient matrix. For a HAADF-STEM image with a shape (512,512), this procedure takes only seconds on a modest desktop. It has the potential to enable real-time symmetry map generation during the collection of experimental STEM images.

%

\begin{figure}[!h]
    \includegraphics[width=\textwidth]{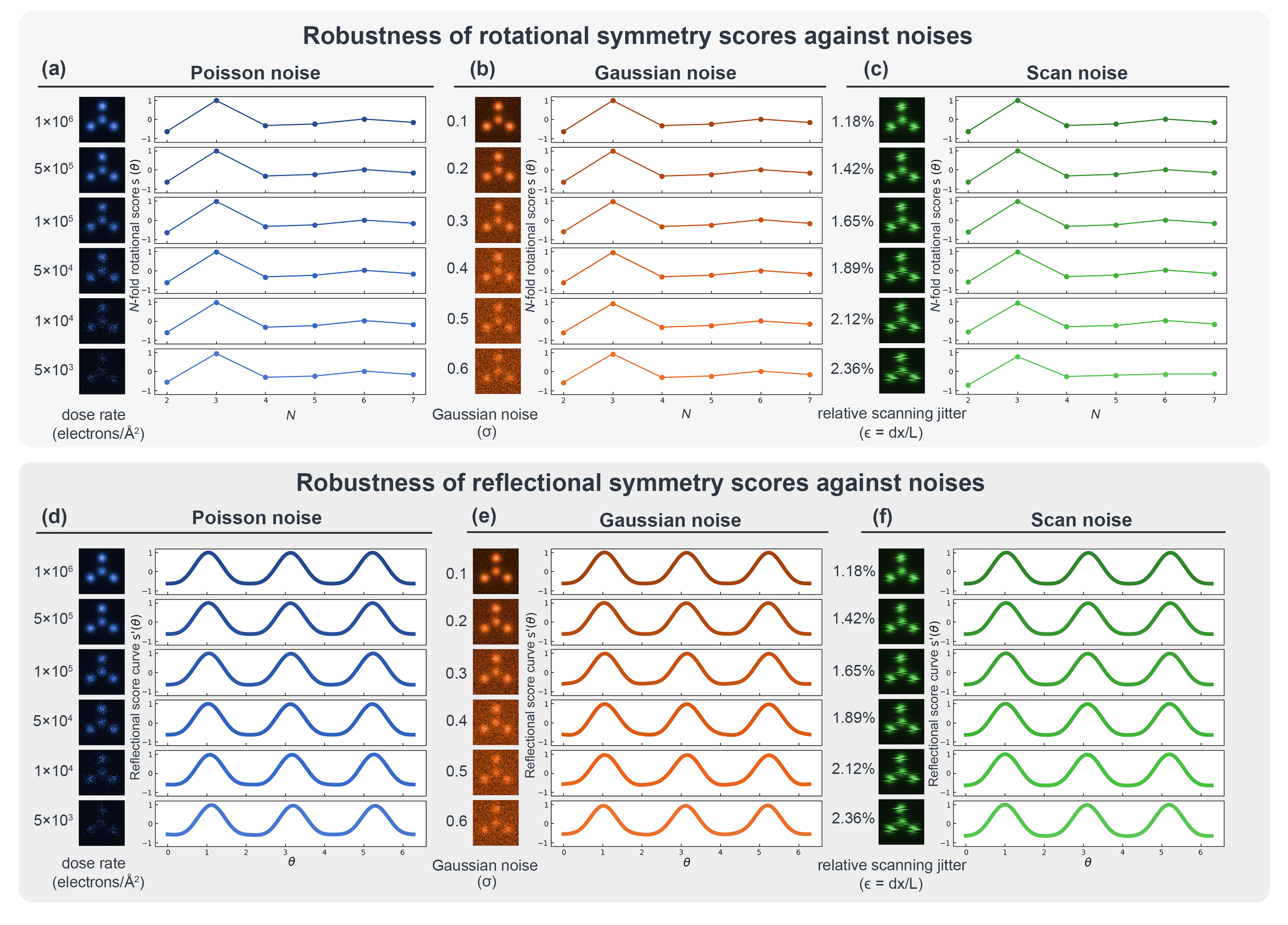}
    \captionsetup{font=small}
    \caption{Robustness of rotational and reflectional symmetry scores against different types of noise. ($\textbf{a}$)--($\textbf{c}$) Rotation symmetry scores $S_N$ on simulated images with different Poisson, Gaussian, and scan noise. ($\textbf{d}$)--($\textbf{f}$) Reflectional symmetry scores $S_R$ on simulated images with different Poisson, Gaussian, and scan noise.}
    \label{fig:Fig3}
\end{figure}

\section{Noise robustness of symmetry measurement scores in STEM imaging }

Our rotational and reflectional symmetry scores demonstrate resilience against different types of noise, making them an ideal universal symmetry descriptor. This resilience is particularly valuable in applications like low-dose imaging\ucite{25,26} and quantum defect identification\ucite{19,20}, where data often contains substantial unwanted variations or disturbances. We specifically select Poisson, Gaussian, and scan noise models for our evaluation due to their prevalence in experimental STEM images, allowing us to comprehensively assess the resilience of our symmetry measurement scores ($S_N$ and $S_R$). 

We use the multislice algorithm\ucite{27} to simulate a monolayer MoSe\textsubscript{2} (see Appendix \hyperref[appen:A1]{A1} for simulation details). We select the Mo-centered motif as our clean reference patch. To simulate real-world conditions, we apply Poisson noise to the reference patch, generating a spectrum of noisy patches with effective dose rates ranging from $10^6$ to $5\times10^3$ electrons$/$\AA$^2$ (depicted in blue colormap in Fig. \hyperref[fig:Fig3]{\ref{fig:Fig3}}). For the Gaussian noise scenario, we add Gaussian noise of different $\sigma$ values ($0.1<\sigma<0.6$) to our clean image, resulting in a series of noisy images in orange. In the case of scan noise, we apply horizontal distortions ($1.18\%<\epsilon<2.36\%$) to the clean reference, producing a set of distorted images in green as illustrated in Fig. \hyperref[fig:Fig3]{\ref{fig:Fig3}}.

For each noisy patch, we compute rotational scores from $S_2$ to $S_7$ and visually present these scores alongside the patch itself (Figs. \hyperref[fig:Fig3]{\ref{fig:Fig3}(a)}--\hyperref[fig:Fig3]{\ref{fig:Fig3}(c)}). Throughout the three different levels of noise, an increase in noise level, consistently results in a notable rise in the $S_3$ score across all observations, aligning with expectations due to the Mo-centered motif's inherent 3-fold rotational symmetry. Additionally, the fluctuation in $S_N$ scores remains minimal.

Moreover, we assess the reflection curves for each patch to evaluate the reflectional symmetry score's robustness across various noise levels (Figs. \hyperref[fig:Fig3]{\ref{fig:Fig3}(d)}--\hyperref[fig:Fig3]{\ref{fig:Fig3}(f)}). Despite the increase of the noise, the reflection curves across the three noise conditions maintain a similar pattern, each showcasing three prominent peaks. These peaks accurately reflect the orientations of the Mo-centered motif's three mirror planes, further confirming the motif's 3-fold rotational symmetry.

%
\begin{figure}[!h]
    \includegraphics[width=\textwidth]{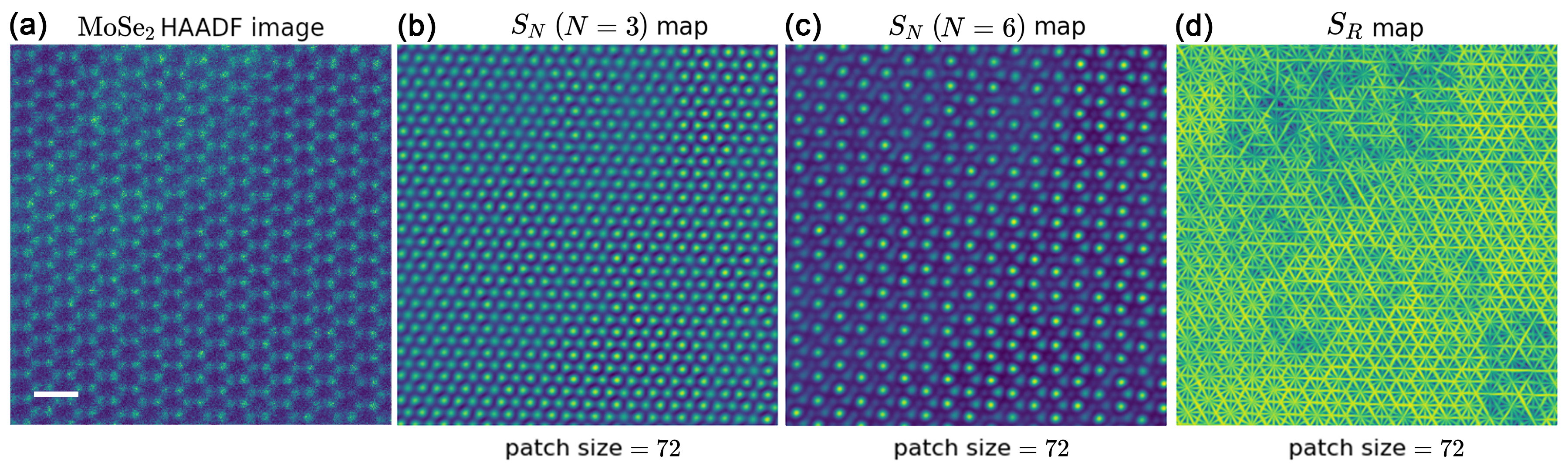}
    \captionsetup{font=small}
    \caption{Rotational and reflectional symmetry maps of an experimental monolayer MoSe\textsubscript{2} HAADF image. ($\textbf{a}$) depicts the experimental HAADF-STEM image of a monolayer MoSe\textsubscript{2} with defects and contamination (upper left). ($\textbf{b}$)--($\textbf{d}$) 3-fold, 6-fold, and reflectional symmetry maps calculated from the experimental image in panel a. The lattice constant $a$ is approximately 27.7 pixels. The patch size used to compute the rotational symmetry and reflectional symmetry is 72 pixels, which approximately corresponds to $\frac{3\sqrt{3}}{2}a$. (scale bar: 0.5 nm)}
    \label{fig:map}
\end{figure}

\section{Rotational and reflectional symmetry maps}

The rotational symmetry and reflectional symmetry maps of an experimental image are shown in Fig. \hyperref[fig:map]{\ref{fig:map}}. In Fig. \hyperref[fig:map]{\ref{fig:map}(a)}, we present an experimental HAAD-STEM image of a monolayer MoSe\textsubscript{2}. Notably, this image exhibits multiple defect points and contamination primarily localized in the upper-left area. The resulting 3-fold and 6-fold rotational symmetry maps are shown in Figs. \hyperref[fig:map]{\ref{fig:map}(b)} and \hyperref[fig:map]{\ref{fig:map}(c)}, and the reflectional symmetry map is illustrated in Fig. \hyperref[fig:map]{\ref{fig:map}(d)}. The patch size is a very important hyperparameter to compute the rotational and reflectional symmetry maps. We estimate the patch size based on the first coordination shell radius of atomic columns. For MoSe\textsubscript{2}, the patch size is 72 pixels, equivalent to $\frac{3\sqrt{3}}{2}a$, where $a$ is the lattice constant. The details are discussed in Appendix \hyperref[appen:A2]{A2}.

The rotational symmetry maps demonstrate minimal susceptibility to point defects, noise, and contamination. In Figs. \hyperref[fig:map]{\ref{fig:map}(b)} and \hyperref[fig:map]{\ref{fig:map}(c)}, there is no obvious difference between upper-left and other areas caused by contamination in the rotational symmetry maps. High score values of 3-fold and 6-fold rotational symmetry observed at atomic positions align seamlessly with the underlying symmetry points, reinforcing the fidelity of the symmetry mapping. 

The reflectional symmetry map as depicted in Fig. \hyperref[fig:map]{\ref{fig:map}(d)} also demonstrates a notable tolerance to noise and contamination, however, it reveals a higher sensitivity to structural variations (\eg, point defects in our cases) compared to its rotational symmetry map. Distinct mirror planes are observed in our symmetry map although the defects impair the regularity of mirror planes. This distinctive sensitivity empowers the use of reflectional symmetry maps as a useful tool for discerning subtle structure variation in atomic resolution imaging.

%

\begin{figure}[!h]
    \includegraphics[width=\textwidth]{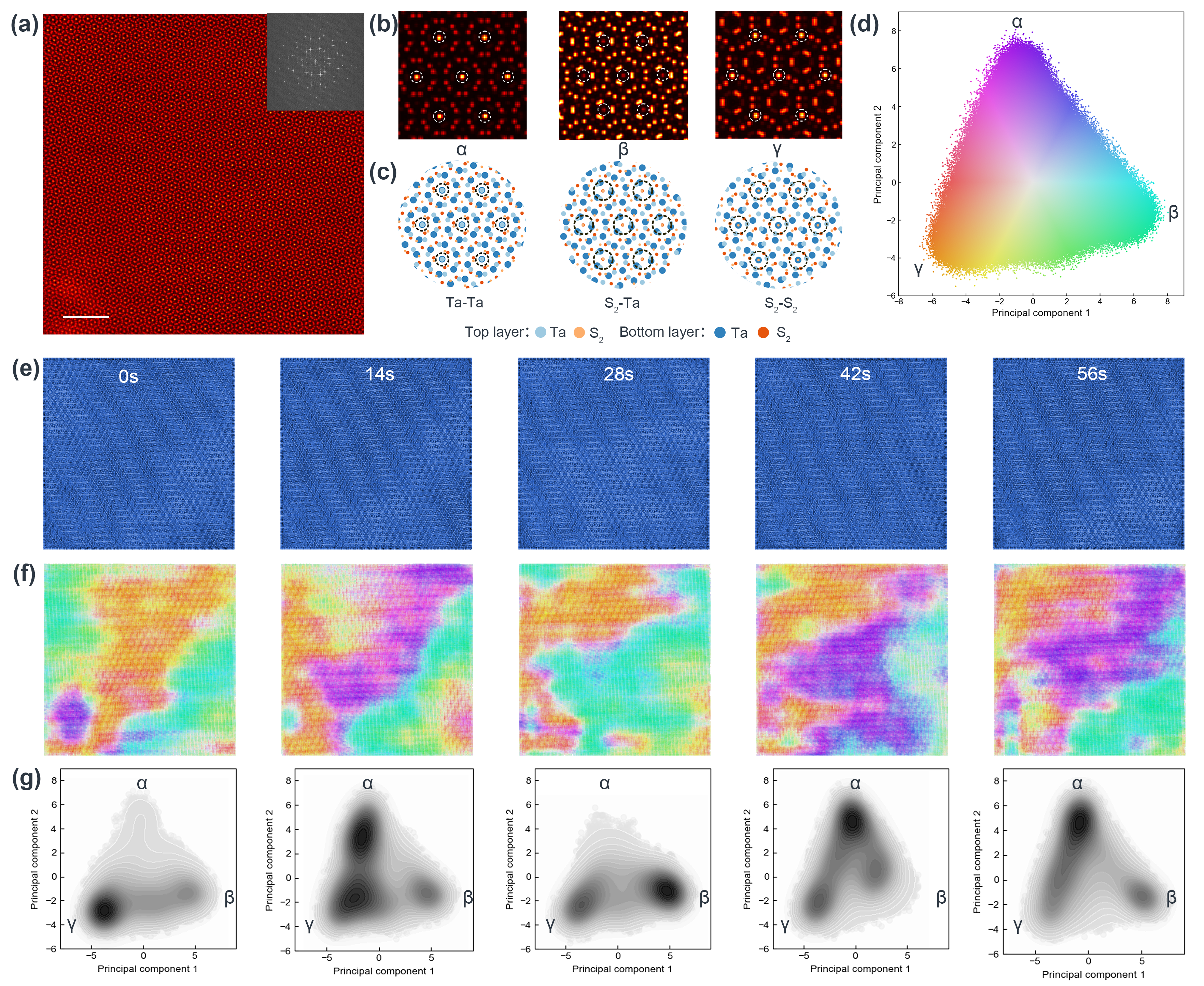}
    \captionsetup{font=small}
    \caption{Reflectional symmetry maps sensitively delineate phases in a twisted bilayer TaS\textsubscript{2}. (\textbf{a}) Depicts the experimental HAADF-STEM image and Fast Fourier Transform image (power spectrum) of the twisted bilayers TaS\textsubscript{2}. (\textbf{b}) shows multislice simulation images of three stacking phases: $\alpha$-TaS\textsubscript{2}, $\beta$-TaS\textsubscript{2}, and $\gamma$-TaS\textsubscript{2}. (\textbf{c}) illustrates the corresponding atomic structures of the three different phases in panel b. (\textbf{d}) displays the 2D PCA layout of local FFT patterns of all reflectional symmetry maps in panel e. (\textbf{e}) represents the reflectional symmetry map of HAADF-STEM twisted bilayer TaS\textsubscript{2} images across various time series. (\textbf{f}) demonstrates the phase segmentation results based on the reflectional symmetry maps in panel e. (\textbf{g}) provides density plots of phases in 2D PCA space as shown in panel e, and these density plots share the same principal components as in panel d. (scale bar: 4 nm)}
    \label{fig:Fig4}
\end{figure}

\section{Unsupervised segmentation of polytypes in twisted bilayer TaS\textsubscript{2}}

By taking a sliding window to extract patches from an experimental STEM image, we can calculate $S_R$ patch by patch to form a reflectional symmetry map. Generally, we find the reflectional symmetry maps generated through our method exhibit high sensitivity to subtle structure variations. This sensitivity makes them exceptionally effective for delineating structural phases within crystalline samples. Specifically, Fig. \hyperref[fig:Fig4]{\ref{fig:Fig4}} highlights the application of this technique on HAADF-STEM images of a twisted bilayer of TaS\textsubscript{2}, showcasing the discernible segmentation of various stacking polytypes.

Figure \hyperref[fig:Fig4]{\ref{fig:Fig4}(a)} shows an experimental HAADF-STEM image of a $\Sigma=7$ bilayer TaS\textsubscript{2} with a twist angle of 21.8$\degree$. Due to the slight displacement between the two layers, the coincidence site lattices of the top and bottom layers have three distinct alignments: Ta--Ta, Ta--S\textsubscript{2}, and S\textsubscript{2}--S\textsubscript{2}. These three stacking variants are also denoted as $\alpha$-TaS\textsubscript{2}, $\beta$-TaS\textsubscript{2}, and $\gamma$-TaS\textsubscript{2} respectively. Fig. \hyperref[fig:Fig4]{\ref{fig:Fig4}(b)} illustrates the multislice simulations of these three phases. Additionally, Fig. \hyperref[fig:Fig4]{\ref{fig:Fig4}(c)} shows the corresponding atomic models for each phase, providing a comprehensive visual understanding of these structural variations. 

To elucidate transitions between these stacking polytypes, which can be initiated by electron beam irradiation\ucite{28}, a series of HAADF-STEM images were captured over time from the same sample region. Given the challenges of directly observing these transitions in experimental images, we developed a novel approach to achieve segmentation of the structural phases. This involves transforming the images into reflectional maps (Fig. \hyperref[fig:Fig4]{\ref{fig:Fig4}(e)}) and employing a sliding window technique to obtain a series of local patches. We use Fast Fourier Transform (FFT) to convert these patches into stacks of power spectra. The local FFT patterns (\ie, power spectra), considered as translation invariant features, undergo Principal Component Analysis (PCA) to be projected into its first two principal components (PC1 and PC2), as illustrated in Fig \hyperref[fig:Fig4]{\ref{fig:Fig4}(d)}. By assigning unique colors to each feature within the PCA space and mapping these colors back to the image space, we generate detailed segmentation maps (Fig. \hyperref[fig:Fig4]{\ref{fig:Fig4}(f)}). These maps, along with density plots (Fig. \hyperref[fig:Fig4]{\ref{fig:Fig4}(g)}), effectively demonstrate the dominance of each phase in the observed frames, providing a clearer perspective on structural dynamics in twisted bilayer TaS\textsubscript{2}. The interfaces between $\alpha$-TaS\textsubscript{2}, $\beta$-TaS\textsubscript{2}, and $\gamma$-TaS\textsubscript{2} are not gradual, as shown by the density plots in the PCA layouts in Fig. \hyperref[fig:Fig4]{\ref{fig:Fig4}(g)}. If the phases transitioned gradually, we would see a relatively uniform density distribution. Instead, we observe non-uniform densities (\ie, dark blobs in density plots), indicating the distinct presence of different phases in different frames.

\section{Discussion and Conclusion}

In this study, we have introduced a methodological advancement for quantifying rotational and reflectional symmetries in image patches through Zernike moments, tailored specifically for STEM imaging applications. Our approach eliminates the need for traditional pixel-based manipulation---such as rotating or reflecting the image itself---which can be computationally demanding. Instead, Zernike moments allow us to abstract these symmetries directly from the image data in a more computationally efficient manner, offering a robust alternative for symmetry assessment. This robustness is particularly critical in the context of STEM imaging, where noise resilience enhances the utility of our method for low-dose imaging\ucite{25,26} and the identification of quantum defects\ucite{29} and structural disorder\ucite{30,31}.

Furthermore, the practical application of our method in delineating structural phases within twisted bilayer TaS\textsubscript{2} showcases its potential to impact materials science. By facilitating the segmentation of structural polytypes and monitoring their transitions under electron beam irradiation, our method offers a powerful tool for the investigation of materials at the atomic scale.

Conclusively, our work not only presents a step forward in the computational analysis of material symmetries but also sets the stage for future innovations in material characterization. By demonstrating how our methods can dissect complex symmetry patterns and reveal underlying structural behaviors in materials, we illuminate paths for breakthroughs in understanding materials at an atomic. Such insights have the potential to catalyze the development of innovative materials with optimized properties, possibly contributing to the evolution of next-generation technologies in energy, electronics, and beyond.

\section*{Appendix A}

\subsection*{A1. STEM characterization and multislice simulation} \label{appen:A1}

Atomic-resolution STEM-ADF imaging was performed on an aberration-corrected ARM200F,
equipped with a cold field-emission gun and an ASCOR corrector operating at 80 kV. The HAADF-STEM images were collected using a half-angle range from 81 to 280 mrad. The convergence semiangle of the probe was 30 mrad. In the simulation, the energy of a convergent electron beam is 80 keV and the cutoff convergence semiangle is 30 mrad. We use the abTEM package (Ref. \cite{27}) to conduct the multislice simulations.

\subsection*{A2. Estimation of patch size for Zernike polynomials} \label{appen:A2}

The patch size to compute rotational and reflectional symmetry maps is related to some characteristic lengths, such as the first coordination shell radius, which can be estimated from the experimental image. Note that the characteristic length in most cases is not equal to lattice constant $a$. However, for convenience, we put lattice constant $a$ here as a reference as it is easier to relate to lattice structure. For rotational symmetry maps, the minimum patch size should be $\sqrt{3}a$ to include the first nearest neighbor atomic columns. For the reflectional symmetry map, the minimum patch size should be $\frac{3}{2}\sqrt{3}a$ to include the second and third nearest neighbors. The lattice constant $a$ can be estimated in the Fourier space\ucite{19}, and it follows three steps: (1) compute the power spectrum of the experimental image; (2) obtain the radial average of the power spectrum image; (3) estimate characteristic length from the first peak of the radial average curve.

\subsection*{A3. Calculation of coefficients $w_{iN}$ in rotational symmetry score} \label{appen:A3}

In Eqs. (\ref{eq:SN}) and (\ref{eq:wiN}), the pre-determined coefficient $w_{iN}$ is defined as $ \frac{1}{N-1}\sum_{k = 1}^{N-1} \cos{(m \frac{2 \pi k}{N} )}$, which varies based on the values of $m$ and $N$. This coefficient is characterized by three distinct cases: $m \leq 1$; $m \geq 2 \land m\pmod{N} = 0$; and $m \geq 2 \land m\pmod{N} \neq 0$.

For case (1), in practice, Zernike moments with $m=0$ capture the radial symmetry of an image, and Zernike moments with $m=1$ are related to inversion and reflection symmetries. In both cases, while they contribute to the overall representation of an image in terms of its general shape and size, they do not provide information on angular features that define \textit{N}-fold rotational symmetry. Therefore, we set the coefficients $w_{iN}=0$ for $m=0$ and $m=1$. 

For case (2), when $m \geq 2 \land m\pmod{N} = 0$, we assume $m = aN$, and a is an positive integer:
\begin{align*}
w_{iN} &= \frac{1}{N-1}\cdot \sum_{k = 1}^{N-1} \cos{(m \frac{2 \pi k}{N} )} \\
&= \frac{1}{N-1}\cdot \sum_{k = 1}^{N-1} \cos{(a N \frac{2 \pi k}{N} )} \\
&= \frac{1}{N-1}\cdot \sum_{k = 1}^{N-1} \cos{(2\pi k  \frac{aN}{N})}\\
&= \frac{1}{N-1}\cdot \sum_{k = 1}^{N-1} \cos{(2\pi k a)}\\
&= \frac{1}{N-1}\cdot \sum_{k = 1}^{N-1} 1\\
&= \frac{1}{N-1}\cdot \left(N-1\right)\\
&= 1
\end{align*}

For case (3), when $m \geq 2 \land m\pmod{N} \neq 0$,  $\rm{j}$ is defined as the imaginary unit, which satisfies $\rm{j}^2=-1$:
\begin{align*}
w_{iN} &= \frac{1}{N-1}\cdot \sum_{k = 1}^{N-1} \cos{(m \frac{2 \pi k}{N} )}\\
&=\frac{1}{N-1}\cdot \sum_{k = 1}^{N-1} \frac{1}{2}\cdot (e^{\frac{2\pi m k}{N}\rm{j}}+e^{-\frac{2\pi m k}{N}\rm{j}})\\
&=\frac{1}{N-1}\cdot \frac{1}{2}\cdot \left(\sum_{k = 1}^{N-1} e^{\frac{2\pi m k}{N}\rm{j}} + \sum_{k = 1}^{N-1} e^{-\frac{2\pi m k}{N}\rm{j}}\right)\\
&= \frac{1}{N-1}\cdot \frac{1}{2}\cdot \left( \frac{e^{2\pi \rm{j} m} - e^{\frac{\left(2 \pi \rm{j} m\right)}{N}}}{-1+e^{\frac{\left(2 \pi \rm{j} m\right)}{N}}}-\frac{1-e^{-\frac{\left(2 \pi \rm{j} \left(N-1\right) m \right)}{N}}}{1-e^{\frac{2 \pi \rm{j} m}{N}}}  \right)\\
\end{align*} 
Since $m$ is integer, $e^{2\rm{j}\pi m} = 1$. So we have:
\begin{align*}
w_{iN}&=\frac{1}{N-1}\cdot \frac{1}{2}\cdot \left(\frac{1 - e^{\frac{\left(2 \rm{j} \pi m\right)}{N}}}{-1+e^{\frac{\left(2 \rm{j} \pi m\right)}{N}}} -\frac{1-e^{-2 \rm{j} \pi m \left(1- \frac{1}{N}\right)}}{1-e^{\frac{2 \rm{j} \pi m}{N}}} \right)\\
&= \frac{1}{N-1}\cdot \frac{1}{2} \cdot \left(\frac{-1\left(-1 + e^{\frac{\left(2 \rm{j} \pi m\right)}{N}}\right)}{-1+e^{\frac{\left(2 \rm{j} \pi m\right)}{N}}}  -  \frac{1-e^{-2 \pi \rm{j} m}\cdot e^{\frac{2 \pi \rm{j} m}{N}}}{1-e^{\frac{2 \pi \rm{j} m}{N}}}  \right)\\
&= \frac{1}{N-1}\cdot \frac{1}{2}\cdot \left( -1 - \frac{1-1\cdot e^{\frac{2 \pi \rm{j} m}{N}}}{1-e^{\frac{2 \pi \rm{j} m}{N}}}         \right)\\
&= \frac{1}{N-1}\cdot \frac{1}{2}\cdot \left(-1 - \frac{1-e^{\frac{2 \pi \rm{j} m}{N}}}{1-e^{\frac{2 \pi \rm{j} m}{N}}}  \right)\\
&= \frac{1}{N-1}\cdot \frac{1}{2}\cdot \left( -1 -1 \right)\\
&= \frac{1}{N-1}\cdot \frac{1}{2}\cdot \left( -2 \right)\\
&= - \frac{1}{N-1}
\end{align*} 
Therefore, the pre-determined coefficient is:

\begin{equation*}
w_{i N}= \begin{cases}0, &\text{ when } \quad m \leq 1 \\
1, \quad &\text { when } \quad m \geq 2 \land m\pmod{N}=0 \\
-\frac{1}{N-1},  &\text { when } \quad m \geq 2 \land m\pmod{N} \neq 0\end{cases}
\end{equation*}

%
\section*{Code availability} \label{code-availability}

The source code and algorithms from this study are available on our public GitHub repository, \href{https://github.com/jiadongdan/motif-learn}{motif-learn}. This repository contains all the necessary scripts and tutorial notebooks needed to reproduce our experiments and apply our methods to new projects. 

%
\addcontentsline{toc}{chapter}{Acknowledgment}
\section*{Acknowledgment}

N.D.L. acknowledges funding support from the National Research Foundation (Competitive Research Program grant number NRF-CRP16-2015-05) and the National University of Singapore Early Career Research Award. This project is supported by the Eric and Wendy Schmidt AI in Science Postdoctoral Fellowship, a Schmidt Futures program.

%

\end{CJK*}  
\end{document}